\documentclass[twocolumn]{aastex631}

\newcommand{\lya}{Ly$\alpha$}

\graphicspath{{./}{figures/}}

\usepackage{amsmath, amsfonts, amssymb, bm} 
\usepackage{natbib} 
\usepackage{ulem}

\shorttitle{Fast Lyman $\alpha$ forests}
\shortauthors{Harrington et al.}

\mathchardef\mhyphen="2D 
\newcommand{\Lya}{Ly$\alpha$} 

\begin{document}

\title{Fast, high-fidelity Lyman $\alpha$ forests with convolutional neural networks}

\author{Peter Harrington}
\affiliation{Lawrence Berkeley National Laboratory, 1 Cyclotron Road, Berkeley, CA 94720, USA}

\author{Mustafa Mustafa}
\affiliation{Lawrence Berkeley National Laboratory, 1 Cyclotron Road, Berkeley, CA 94720, USA}

\author{Max Dornfest}
\affiliation{Lawrence Berkeley National Laboratory, 1 Cyclotron Road, Berkeley, CA 94720, USA}

\author{Benjamin Horowitz}
\affil{Department of Astronomy, Princeton University, Princeton, NJ, USA}
\affil{Lawrence Berkeley National Laboratory, 1 Cyclotron Road, Berkeley, CA 94720, USA}

\author{Zarija Luki\'{c}}
\affiliation{Lawrence Berkeley National Laboratory, 1 Cyclotron Road, Berkeley, CA 94720, USA}

\begin{abstract}

Full-physics cosmological simulations are powerful tools for studying the formation and evolution of structure in the universe but require extreme computational resources. Here, we train a convolutional neural network to use a cheaper N-body-only simulation to reconstruct the baryon hydrodynamic variables (density, temperature, and velocity) on scales relevant to the Lyman-$\alpha$ (\Lya{})  forest, using data from \texttt{Nyx} simulations. We show that our method enables rapid estimation of these fields at a resolution of $\sim$20kpc, and captures the statistics of the \Lya{} forest with much greater accuracy than existing approximations. Because our model is fully-convolutional, we can train on smaller simulation boxes and deploy on much larger ones, enabling substantial computational savings. Furthermore, as our method produces an approximation for the hydrodynamic fields instead of \Lya{} flux directly, it is not limited to a particular choice of ionizing background or mean transmitted flux.

\end{abstract}


\section{Introduction} \label{sec:intro}

Understanding the distribution and evolution of the matter in the universe is at the core of modern cosmology and a main motivation for the upcoming generation of astronomical surveys. But to extract cosmological insights, observations have to be compared to theoretical predictions for different cosmological scenarios.  Predicting observable quantities in general, and \Lya{} forest specifically, from the underlying large scale structure is complicated by the nonlinear mapping between baryons (responsible for most of observables) and dark matter (which dominates the gravitational dynamics).  In the early days preceding ``precision cosmology'', simplified models were successfully used, for example baryons-trace-dark-matter approximation \citep{Petitjean1995}, or modifying the matter density field \citep{Gnedin1996} or gravitational potential \citep{Gnedin1998}, to mimic the effects of baryonic pressure smoothing.  The gas temperature of the intergalacic medium (IGM) in those models is usually computed by imposing a polytropic temperature–density relation \citep{Hui1997}.

More recently, recipes to model \Lya{} forest without running full hydrodynamical simulations have been provided by \cite{Peirani2014} and \cite{2016Sorini}, as well as models for specific summary statistics in \cite{2012JCAP...03..004S}, but they have not been really used in practice due to their complexity and/or applicability to only a limited range of scales.  Thus, \Lya{} forest forward models rely on computationally expensive hydrodynamical simulations \citep{Irsic2017, Boera2019, Walther2019, PDB2020, Rogers2020, Walther2021}, and solving inverse problems commonly requires running dozens of simulations, at a rough cost of 2 million CPU hours per simulation \citep[see, for example,][]{Walther2021}. Of course, if one wants to consider cosmological models beyond standard $\Lambda$CDM or to model complex astrophysical phenomena affecting the \lya\ signal like high column density absorbers \citep{Rogers2018}, UV background fluctuations \citep{Onorbe2019}, or HeII reionization \citep{UptonSanderbeck2020} the number of needed simulations, and thus the computational costs, will increase significantly. Such applications present a clear need for surrogate models which can mitigate some of the cost of running full-fledged hydrodynamic simulations for every test case.

Beyond generation of mock catalogs and hydrodynamic emulators, there is also a growing need for fast hydrodynamical mapping in the context of differentiable forward model reconstructions \citep{2019TARDISI}. There, the underlying matter distribution is reconstructed under constraints from gravitational evolution and using, for example, the three dimensional \lya\ tomographic reconstructions under the Fluctuating Gunn-Peterson Approximation (FGPA; \cite{1965GunnPeterson}) as done in \cite{2019TARDISI,2019Porqueres, 2020TARDISII}. Reducing the modeling cost via fast and efficient --- even if approximate --- methods of reproducing the \Lya{} signal is thus of significant interest to the community.  Our paper addresses this issue using data-driven modeling trained on hydrodynamical simulations.

In recent years, deep neural networks (NN) have become a promising method to assist in solving these highly nonlinear problems, by acting as surrogate models (which are differentiable by design) for complex phenomena. In particular, Generative Adversarial Networks (\citealt{2014GAN}; GANs) and U-Nets \citep{UNet} have proven useful in a variety of cosmology tasks that either directly generate synthetic hydrodynamic quantities \citep{2019HIGan}, or reconstruct them from dark matter distributions. The latter approach allows one to run cheaper N-body simulations (or even surrogates for N-body evolution) to generate a dark matter distribution, then rapidly estimate the target hydrodynamic quantity corresponding to the dark matter structure. In the context of the thermal and kinematic Sunyaev-Zel'dovich effects, this approach has been shown to be useful in reconstructing electron density, presssure, and momenta in 3D \citep{2020tSZelectrons} as well as gas pressure in 2D \citep{2019Painting}. These works showcase the success of neural networks in producing hydrodynamic fields with superior statistical fidelity compared to competing classical or semi-analytical methods. While extremely promising, deep learning approaches are data-hungry, and require generation of training data at the desired output resolution (e.g., $\sim$100 kpc in \citealt{2020tSZelectrons}) to achieve maximal model fidelity. It is thus important to maintain a reasonable volume of training data when extending such models to finer-scale hydrodynamic reconstruction, and investigate the data needs in such regimes.

In this work, we use neural networks to reconstruct 3D baryon hydrodynamics from N-body dark matter simulations at $\sim$20 kpc/$h$ resolution, roughly an order of magnitude finer than previous neural hydrodynamic reconstructions. Given a snapshot of dark matter, our model allows rapid estimation of baryon density, temperature, and velocity fields, which can be painted in over an arbitrarily large volume at the given resolution.

The paper is organized as follows.  We first describe the simulation data used to train our model and detail how we compute FGPA estimates and \Lya{} forest quantities in Section \ref{sec:Data}. Then, we describe the model design and training process in Section \ref{sec:Model}. Our results are presented in Section \ref{sec:Results}, where we show sample output from our model and make statistical comparisons against the true hydrodynamic fields. Finally, we present conclusions and discussion in Section \ref{sec:Conclusion}.

\section{Simulations}
\label{sec:Data}

We construct our training and validation datasets from pairs of cosmological simulations run with the \texttt{Nyx} code \citep{Nyx1, Lukic2015}. Each pair of simulations share identical initial conditions and cosmology, with one simulation modeling the full-physics problem (dark matter particle N-body dynamics as well as baryon hydrodynamics), while the other just models the evolution of dark matter particles (N-body-only). In \texttt{Nyx}, the dark matter particles are evolved with a particle-mesh scheme, while the additional baryon hydrodynamics are modeled as an inviscid ideal fluid on a set of Eulerian grids. We neglect physics related to galaxy formation, but this is extremely common approach in \Lya{} cosmological simulations \citep{Irsic2017, Boera2019, Walther2019, Rogers2020, Walther2021, Pedersen2020}.  This is because regions transparent to \Lya{} photons at redhisfts $z \gtrsim 2$ are in low density regions \citep{Lukic2015, McQuinn2016}, and poorly understood physical processes related to galaxy formation only play a minor role in those regions \citep{Kollmeier2006, Desjacques2006}.\footnote{For a recent work done at higher precision and challenging this view, see \cite{Chabanier2020}}

With this paired dataset, we can train our model to learn a mapping from the cheaper N-body-only simulation into the hydrodynamic fields of the full-physics simulation in a supervised fashion. We use one pair of simulations for training, and an independent pair, with identical cosmology but different initial conditions, for validation.  For both the training and validation pairs of simulations, the cosmological parameters are $\Omega_b=0.05$, $\Omega_M=0.31$, $\Omega_L=0.69$, and $h=0.675$. The physical fields of interest are defined on a 3D uniform $1024^3$ grid, spanning a cube of $L= 20$ Mpc/$h$ per side, with periodic boundary conditions. 
For the hydrogen and helium mass abundances we adopted values consistent with the CMB observations and Big Bang nucleosynthesis \citep{Coc2013}: $X_p$=0.76 and $Y_p$=0.24.
Dark matter N-body dynamics are evolved with $1024^3$ particles, then density and velocities are deposited on the grid using Cloud-In-Cell (CIC) interpolation. We select the $z=3$ snapshot in each simulation for training and validation, as it lies in the center of the range of redshifts most relevant for \Lya{} analysis.

Data scales in \texttt{Nyx} files can span many orders of magnitude, so we normalize the data to order unity for training stability. For all velocity fields, we scale linearly to $\mathbf{\hat{v}} = \mathbf{v}  / (9\times 10^7 \, \mathrm{cm}\cdot \mathrm{s}^{-1})$. The baryon density and temperature fields have a large dynamic range, so we use
\begin{align}
    \hat \rho &= \log(\rho/14), \nonumber  \\
    \hat T &= \log (T)/8 - 1.5, \label{eq:normalizations}\\
    \hat \rho_{DM} &= \log (1 + \rho_{DM})/12. \nonumber
\end{align}
where $\rho$ and $T$ are the baryon overdensity and temperature (in Kelvin), respectively, while $\rho_{DM}$ is the overdensity of dark matter.
All network outputs are transformed back into their original units before performing analysis or validation on statistical quantities.

\begin{figure*}
    \centering
    \includegraphics[width=\linewidth]{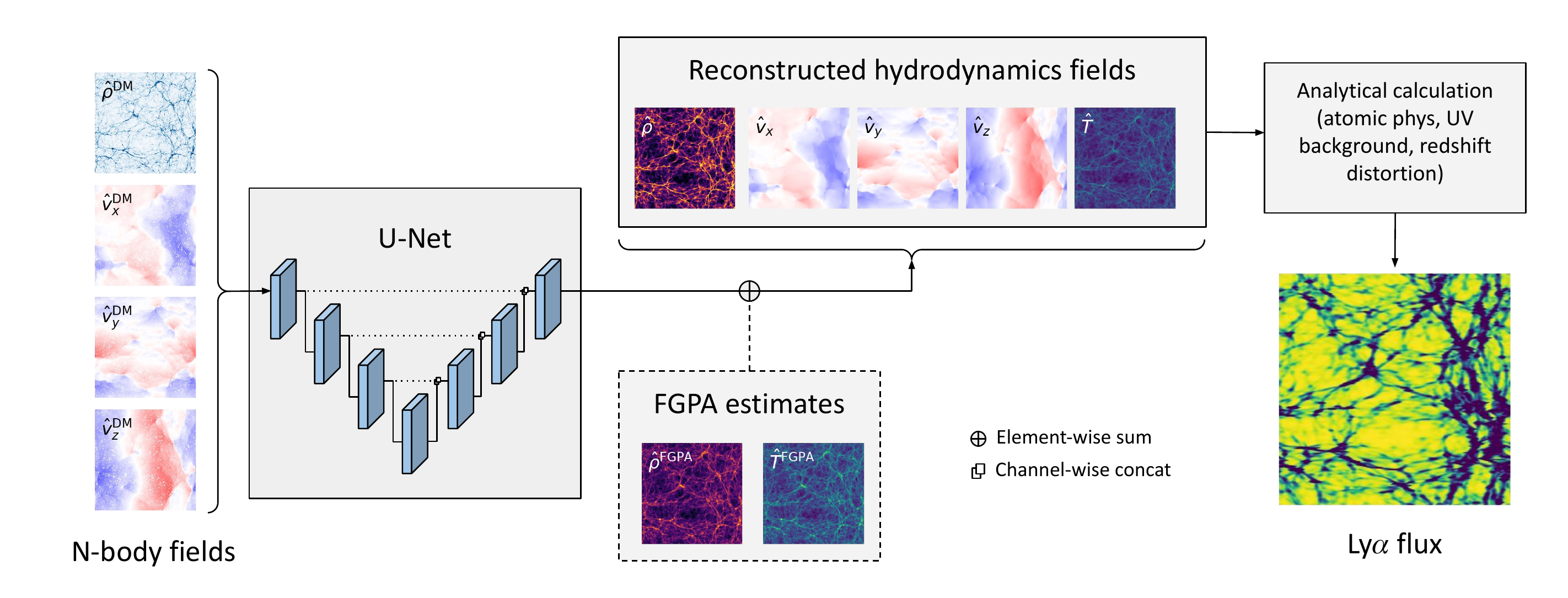}
    \caption{Schematic depicting our model, which predicts the hydrodynamic fields from an input n-body simulation. We train separate networks for $\hat \rho$, $\hat T$, and $\mathbf{\hat{v}} = (\hat v_x, \hat v_y, \hat v_z)$, and for $\hat \rho$ and $\hat T$ we predict the correction to the FGPA approximations $\hat \rho^{\mathrm{FGPA}}, \hat T^{\mathrm{FGPA}}$ rather than the full fields. As the model reconstructs the raw hydrodynamic fields, it is independent of \Lya{}-specific physical details.}
    \label{fig:schematic}
\end{figure*}

\subsection{FGPA estimates}

The long-standing method to reconstruct hydrodynamic quantities for purposes of \Lya{} modeling is the FGPA \citep{1965GunnPeterson, 2016Sorini}). This procedure takes the dark matter overdensity $\rho^{DM}$ from the N-body simulation, with or without artificial smoothing, to produce estimates for the baryon density $\rho$ and temperature $T$, given by a power law relation \citep{Hui1997}
\begin{equation}
    \label{eq:FGPA}
    T = T_0 (\rho/\langle \rho \rangle)^{\gamma -1} \, ,
\end{equation}
where $\langle \rangle$ denotes averaging over volume. The estimate for $\rho$ is usually attained by applying Gaussian smoothing (with smoothing length $\lambda_G$ roughly given by the Jeans filtering scale) to $\rho^{\mathrm{DM}}$ to simulate baryonic pressure smoothing. In this work, we empirically determine the best-fit $T_0$, $\gamma$, and $\lambda_G$ for our simulation.
We best-fit Gaussian smoothing scale using least square differences between the smoothed dark matter density from the N-body simulation and the actual baryon density field from the hydrodynamical simulation.  We then fit density and temperature in the hydrodynamical simulations to the power-law relation in eq.~\ref{eq:FGPA} as described in \cite{Lukic2015}, arriving for our particular simulation and redshift $z=3$, to values:
\begin{align}
    T_0 &= 12,300\, \text{K} \nonumber \\
    \gamma &= 1.49 \\
    \lambda_G &= 43.5\, \text{kpc}/h \nonumber .
\end{align}
We use these values to compute FGPA estimates for the hydrodynamic fields in the test simulation and use the resulting \Lya{} flux as a baseline to compare against the accuracy of our neural network approach. Additionally, FGPA is a useful starting point for the neural network training process, providing an initial guess upon which the network can improve.

\subsection{Lyman $\alpha$ computation}
Given the true or estimated baryon hydrodynamics fields, we can then compute \Lya{} flux analytically. We use the \texttt{Gimlet} code \citep{2016FriesenGimlet} to do this step.
We compute the optical depth, $\tau$, of neutral Hydrogen at a fixed redshift, which is related to the transmitted flux fraction with: $F=\exp{(-\tau)}$. We do not account for the finite speed of light when we cast rays in the simulation, but we use the thermodynamical state of baryons at a fixed cosmic time.  The simulated spectra are thus not meant to fully reproduce observed \lya\ forest spectra, but to recover the flux statistics in a given redshift window. Our calculation of the spectra accounts for Doppler shifts due to bulk flows of the gas and for the thermal broadening of the \lya\ line.  We have neglected noise and metal contamination, but this is both common procedure in the field, and is not relevant for this paper. We refer to \cite{Lukic2015} for specific details of these calculations.  Finally, we rescale the UV background intensity so that the mean flux of all spectra from the network model predictions and hydrodynamical simulation match.
Note that with our neural networks we predict the hydrodynamic fields ($\rho$, $T$, $\bf{v}$) directly, thus we use identical pipelines and assumptions when calculating \lya\ forest quantities from \texttt{Nyx} simulations and NN models.  This makes our approach orthogonal to any particular choices made in modeling \lya\ forest from basic thermodynamical quantities.

\section{Model Design}
\label{sec:Model}

As the target baryon hydrodynamical fields each have their own unique features, we partition the problem and train one generator network each for $\hat{\rho}$, $\hat{T}$, and $\mathbf{\hat{v}}$. We use the same U-Net architecture for our generation networks, as this design is efficient in extracting high-level features through sequential downsampling, while maintaining the ability to resolve fine-grained details via skip connections. The fully-convolutional nature of the U-Net allows us to train on smaller sub-volumes and then predict on larger volumes without issue.

\subsection{Network architecture}

Each downsampling block in our U-Nets consists of a 3D convolution with stride 2, followed by a leaky ReLU activation \citep{LeakyReLU}. There are a total of 6 downsampling blocks, and a corresponding 6 upsampling blocks. Each upsampling block consists of a 3D transposed convolution with stride 1/2, followed by a ReLU activation \citep{ReLU}, except the final upsampling convolution, which has a $\tanh$ nonlinearity. After each upsampling block, the incoming skip connections from the corresponding downsampling block are concatenated to the features along the channel axis.

The input to the network consists of the normalized dark matter density and velocity fields $(\hat{\rho}^{DM}, \hat{v}^{DM}_x, \\ \hat{v}^{DM}_y, \hat{v}^{DM}_z)$, which are supplied as different feature channels over a 3D volume. During training, we generate samples by randomly cropping sub-volumes from the training simulations. The crop size is a tunable hyperparameter, and we find best performance with crops of size $128^3$ for the density and temperature fields, and $256^3$ for the velocity fields. As additional augmentations, we randomly apply rotations and transposes from the octahedral group to these training samples. Together, the random crops and rotations implicitly enforce the large-scale homogeneity and isotropy of the universe, and help extend the utility of our single pair of training simulations.

For the baryon density and temperature targets, we observe a performance benefit when incorporating the FGPA estimates into the training workflow, as depicted in Figure \ref{fig:schematic}. In this setup, the network predictions are added to the FGPA estimate to produce the final field, so the model is trained to predict a correction term to the FGPA estimate rather than the full hydrodynamic field at once. This accelerates convergence and improves the statistical fidelity of our final fields.

\subsection{Loss functions}
The main component of our loss function is the $\mathcal{L}_1$ distance between the generated and target fields. The velocity U-Net is trained entirely with this loss function, but for the baryon density and temperature fields we observe an inability to adequately capture high-density/temperature features training only with $\mathcal{L}_1$ loss. This is consistent with the well-known tendency of $\mathcal{L}_1$ to seek a conservative (maximum-likelihood) estimate and average over fine-scale features, causing noticeable blurring in the network output \citep{2016pix2pix}.

Thus, for the density and temperature fields we also employ an adversarial loss, given by a discriminator $D$. Contrary to a standard adversarial setup, our discriminator operates in Fourier space rather than the original data space. This is motivated by the observation that taking a sample (cube with side length $N$) and removing the $N/2$ smallest-scale Fourier modes along each axis retains much of the important detail, while reducing dimensionality by a roughly an order of magnitude (a factor of 2 along each dimension). With our large 3D training samples, this reduction of dimensionality helps compensate for the added overhead of training an additional discriminator network.

We thus use the following procedure when training our spectral discriminator $D$. The discriminator's input fields $x$ are crops (with side length $N=128$) of the baryon density or temperature field, coming from either the training simulation or our generator network. We compute the discrete Fourier transform of these crops (with wavenumbers $\{ -N/2, ..., 0, ..., N/2 -1\}$) along each dimension, then drop modes with $|k|>= N/4$ to get the truncated Fourier transform  $\Tilde x_t$. Then, we feed the truncated Fourier coefficients $f = \Tilde x_t \Tilde{x}^*_t$ to the spectral discriminator, which tries to classify the sample as real or fake. Our discriminator has four downsampling convolutional layers, followed by three fully-connected layers which end in a single output classifying the input as real or fake. To give the spectral discriminator additional context, we concatenate the dark matter density input as an additional feature channel before applying the Fourier truncation. 

Following the standard non-saturating GAN formulation \citep{2014GAN}, the adversarial loss for the discriminator is 
\begin{equation}
    \mathcal{L}_{D} = -\log D(f_{\rm{real}}) - \log(1 - D(f_{\rm{fake}})),
\end{equation}
while the total loss for the $\hat \rho$ and $\hat T$ U-Nets is
\begin{equation}
    \mathcal{L}_{\mathrm{U\mhyphen Net}} = \lambda \mathcal{L}_1 + \log D(f_{\rm{fake}}),
\end{equation}
where $\lambda=500$ is a hyperparameter to up-weight the importance of the $\mathcal{L}_1$ loss.

\section{Results}
\label{sec:Results}

We evaluate the accuracy of our approach primarily by comparing the PDF \citep{Rauch1997} and 1D power spectra \citep{Croft1999} of our model's \Lya{} fields against those of the test simulation. The \Lya{} flux variation along some line-of-sight axis is defined with respect to fluctuations about the global mean $\bar F$, as $\delta_F = (F - \bar F)/\bar F$. From this, the 1D power spectrum $P(k)$ is given by
\begin{equation}
    P(k) = \langle \widetilde{\delta}_F(k) \widetilde{\delta}^*_F (k) \rangle,
\end{equation}
where $\widetilde{\delta}_F (k)$ is the Fourier transform of $\delta_F$ and the mean $\langle \rangle$ is taken over all modes with magnitude $k$ along the given line-of-sight axis (one flux skewer per spatial location $i,j$). From this we can compute the transfer function $T(k)$, defined by
\begin{equation}
    T(k) = \sqrt{\frac{P^{\mathrm{U\mbox{-}Net}} (k)}{{P^{\mathrm{True}}(k)}}},
\end{equation}
to assess how closely our predicted \Lya{} fields match the test simulation in power spectra. 

During training, candidate models are selected based on how closely their hydrodynamic reconstructions for $\hat \rho$, $\hat T$, and $\mathbf{\hat v}$ match the PDFs of the ground truth fields. Then, the hydrodynamic reconstruction is computed for the entire test simulation at once by applying our network to the full 1024$^3$ volume\footnote{This step is done on CPU rather than GPU due to the memory requirement, but only takes several minutes to complete. Alternatively, one can reduce memory requirements by iteratively predicting on smaller sub-volumes and stitching them together \textit{post hoc} by averaging overlapping edges. We have confirmed this approach does not noticeably degrade the quality of results.}, using periodic padding to enforce the periodic boundary conditions. We choose our final model based on the accuracy of the PDF and $P(k)$ for the \Lya{} fields computed from the hydrodynamic reconstruction. In the following sections we will visualize our \Lya{} and corresponding hydrodynamic fields, and detail how closely they mach the test simulation in the summary statistics. We compare against the FGPA \Lya{} estimates as a baseline. Then, we will analyze the hydrodynamic reconstructions and examine some failure modes.

\subsection{Lyman $\alpha$ fields}

\begin{figure*}
    \centering
    \includegraphics[width=\linewidth]{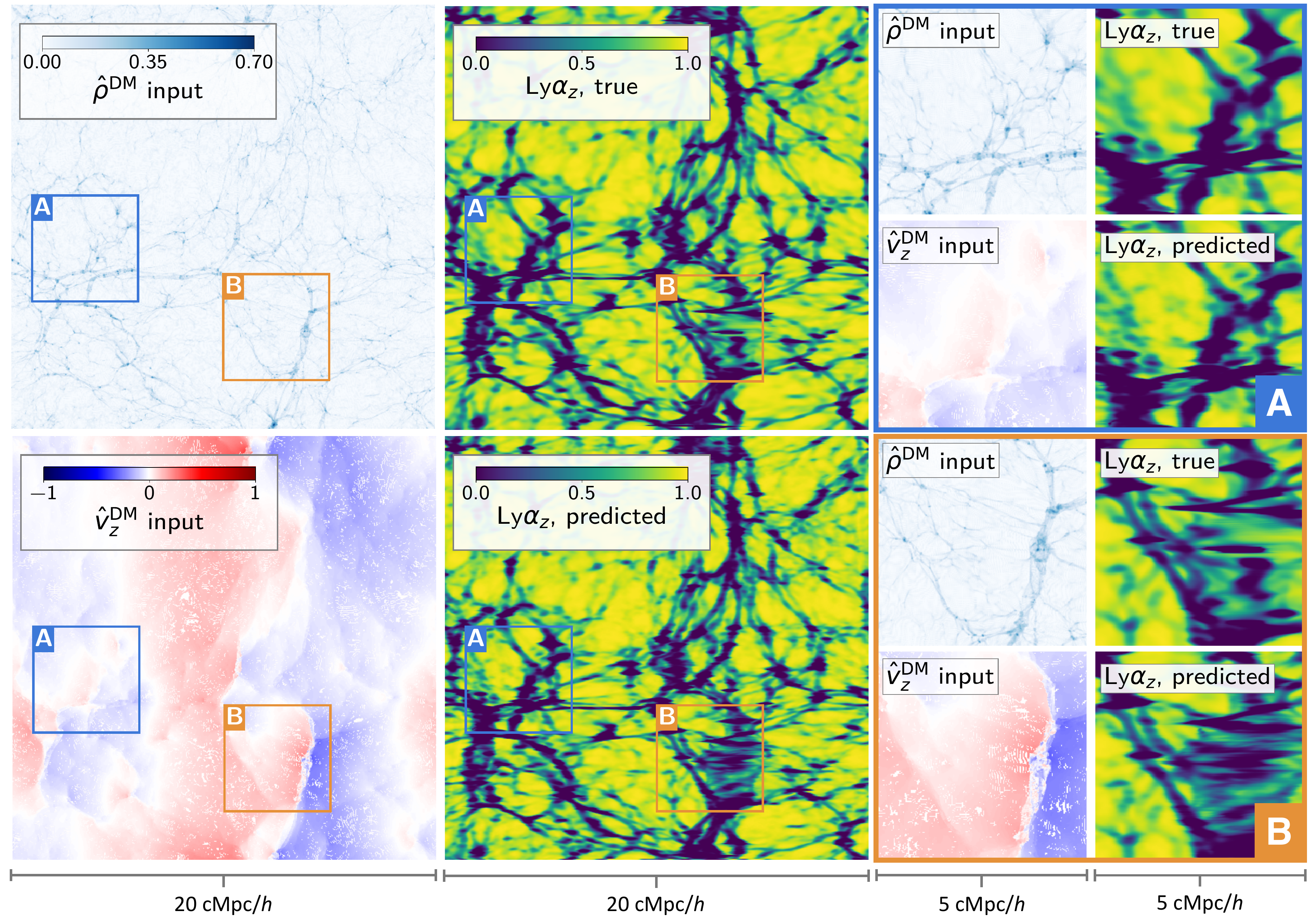}
    \caption{Slices of the test simulation, with the true and predicted \Lya{} fields along the $z$-axis line-of-sight. At left, we show a slice through the full volume of the input dark matter density and the line-of-sight dark matter velocity $v_z$ for reference. The corresponding true and predicted Ly$\alpha_z$ fields for this slice are shown in the center panels, and at left we show zoom-ins to regions A and B to better contrast our model against the ground truth.}
    \label{fig:Lya_fields}
\end{figure*}

A visualization of our results is shown in Figure \ref{fig:Lya_fields}, where we render some of the input fields and corresponding \Lya{} fields at a randomly chosen slice in the $x$-$z$ plane ($z$ was chosen as the line-of-sight here). The slices run the entire length of the simulation box, but several zoomed-in regions are also shown for increased detail on the right-hand side. 

The predicted \Lya{} field has a high degree of visual fidelity, and closely matches the ground truth in diffuse IGM regions where densities are lower or where there is not much gas motion (see zoom-in panel A). By contrast, regions where gas is collapsing onto a filament structure show more of a discrepancy (zoom-in panel B), as the redshift distortion from the velocity field effectively stretches the size of the filament in the redshifted \Lya{} field and amplifies any discrepancies between the true and reconstructed hydrodynamic fields along the filament. In general, dense filaments and clusters make up an exceedingly small fraction of the total volume, so the model has more difficulty capturing the fine-scale features in these regions due to limited training examples.

\begin{figure*}
    \centering
    \includegraphics[width=0.48\linewidth]{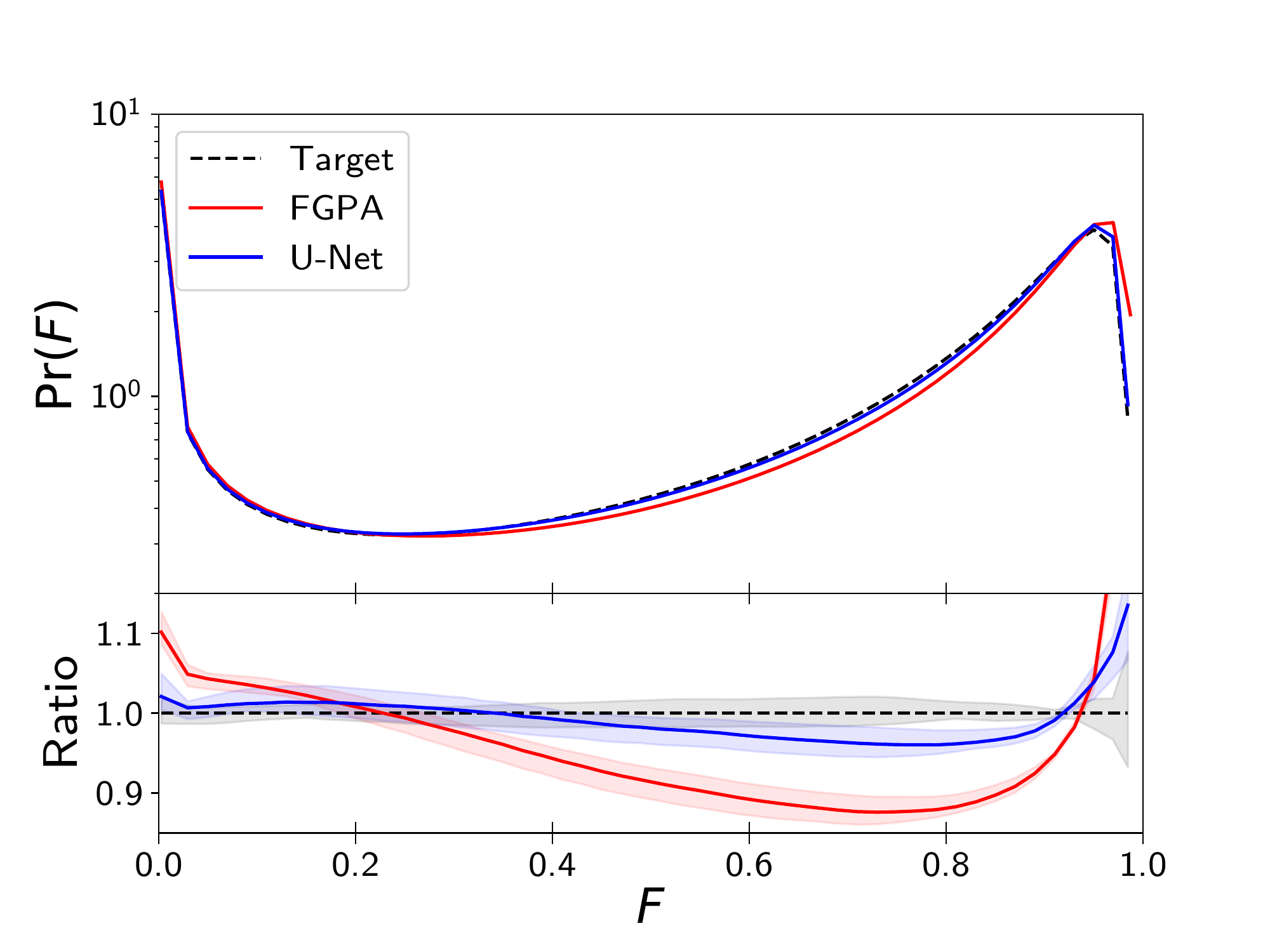}
    \includegraphics[width=0.48\linewidth]{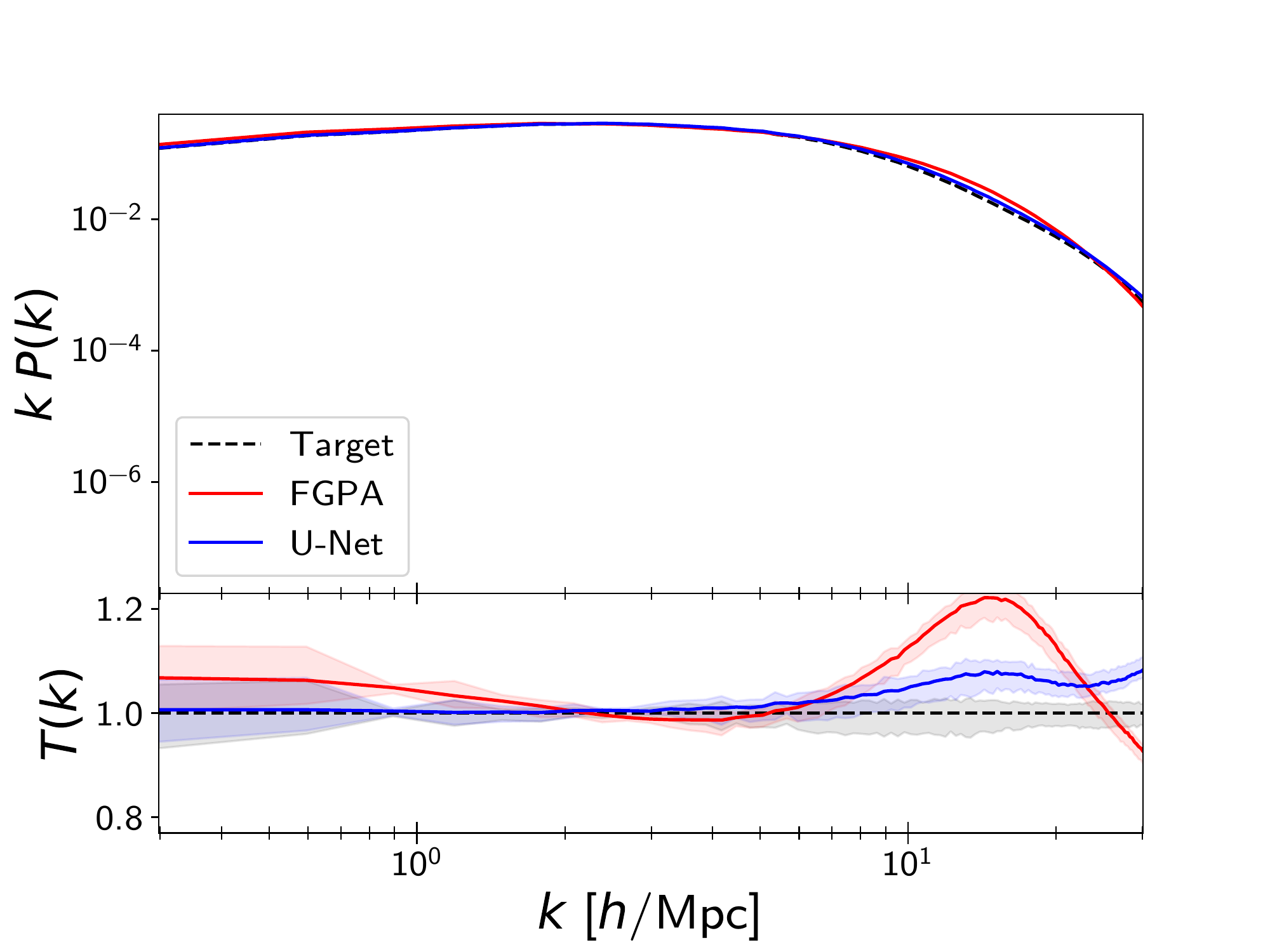}
    \caption{The PDF (left) and power spectrum (right) of full-volume \Lya{} flux fields from our model and FGPA, compared against the test simulation. Lines and shaded regions represent the mean and min/max per bin across \Lya{} fields with lines of sight along the $x$, $y$, and $z$ axes.}
    \label{fig:Lya_stats}
\end{figure*}

Since at $z=3$ the \Lya{} transmission drops to zero at overdensity of a few (see Fig.~7 in \cite{Lukic2015}), we can still obtain statistically accurate \Lya{} fields from our hydrodynamic reconstructions without necessarily capturing the full range of scales in the gas physics. This is demonstrated in Figure \ref{fig:Lya_stats}, where we show the PDF and 1D power spectrum of our \Lya{} estimates and compare against the ground truth. We compute the metrics for \Lya{} fields with lines-of-sight along all three axes of our box, and plot the mean, minimum and maximum per $F$ and $k$ bin across the three fields. 

\begin{figure}
    \centering
    \includegraphics[width=\linewidth]{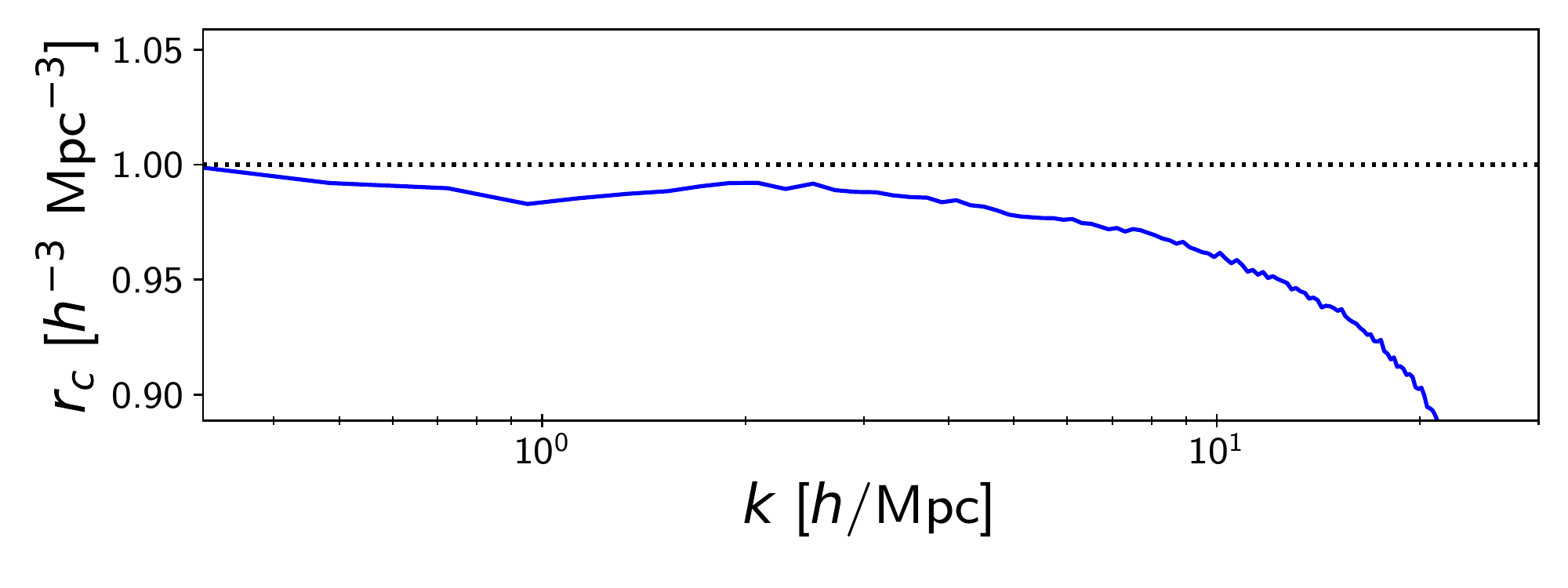}
    \caption{The cross-correlation coefficient $r_c$ of our predicted \Lya{} field over the relevant length scales. We used the $z$-axis line-of-sight for this plot.}
    \label{fig:Lya_CC}
\end{figure}

In both statistical measures, the \Lya{} estimates from our model are within a few percent of the ground truth across the relevant $F$ and $k$ bins, significantly improving on the FGPA \Lya{} estimates. Our model begins to accumulate an excess of power in $P(k)$ near $k\sim10$ $h$/Mpc, as does FGPA. Beyond 10 $h$/Mpc, both our model and FGPA begin to diverge from the truth, with our model having an excess of power at high wavenumbers while FGPA diverges in the other direction. However, $k\sim10$ $h$/Mpc is typically the highest mode considered in \Lya{} spectroscopic studies, limited by contamination from metal lines (see \citealt{2016Sorini} and references therein), so we truncate the plot near that wavenumber and do not consider the discrepancies beyond it to be detrimental to our approach. 

To further evaluate our model, we compute the cross correlation coefficient $r_c$ , defined as 
\begin{equation}
    r_c = \frac{P_{rt}}{\sqrt{P_{rr}P_{tt}}},
\end{equation}
where $P_{rt}$ is the cross-power and $P_{rr}, P_{tt}$ are the auto-power of the reconstructed and true \Lya{} fields, respectively. The cross-correlation for the $z$-axis line-of-sight is shown in Figure \ref{fig:Lya_CC}. We see good agreement, with $r_c>0.95$ across the relevant $k$ range.


\subsection{Hydrodynamic fields}

\begin{figure*}
    \centering
    \includegraphics[width=\linewidth]{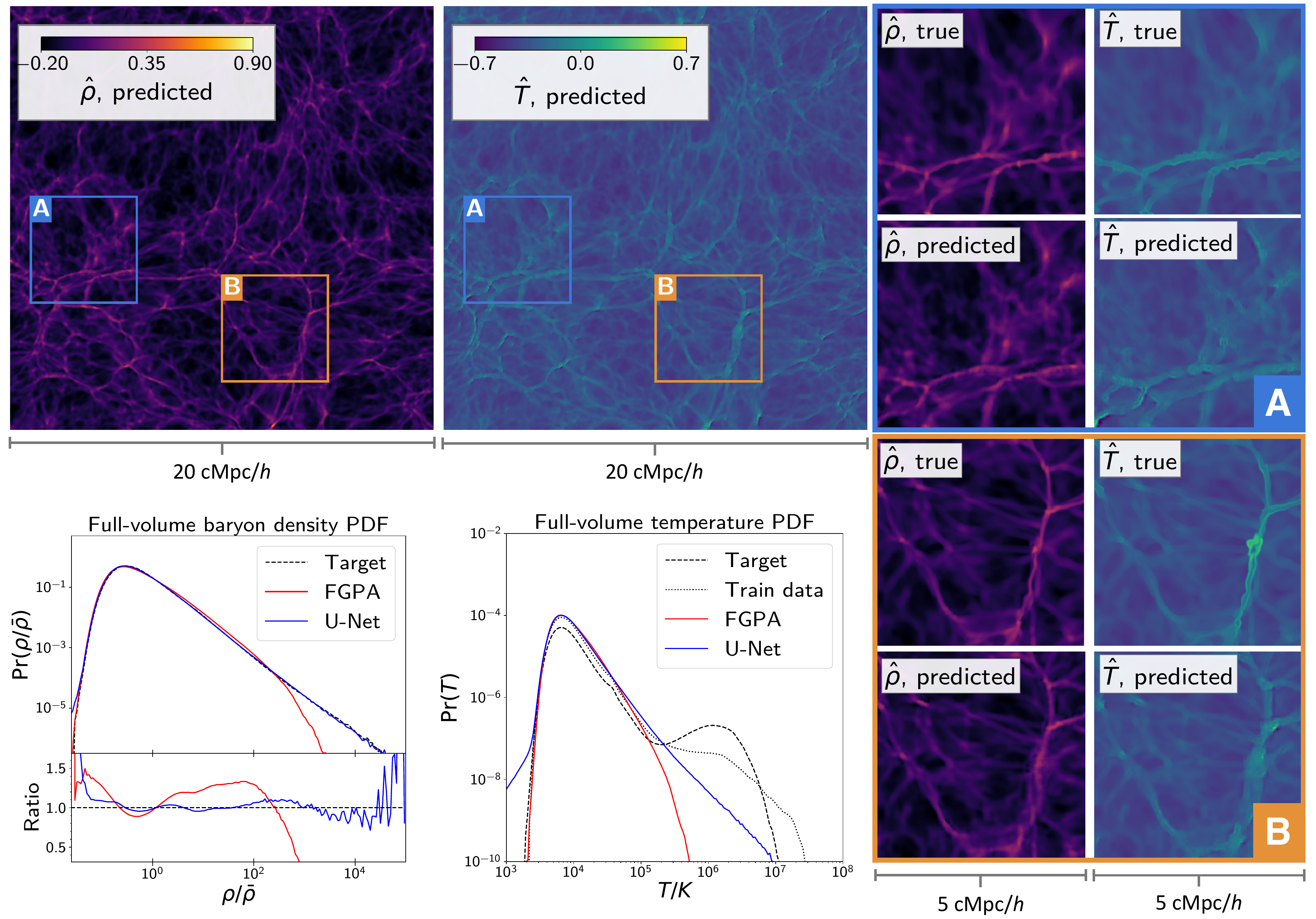}
    \caption{Predicted baryon density and temperature fields, sliced at the same location as plotted in Figure \ref{fig:Lya_fields}. In the right-side panels we show zoom-ins to regions A and B as before. In the lower-left panels, we plot the full-volume PDFs for these fields (after transforming back to original simulation units) compared to the ground truth and FGPA. The reconstructed baryon density field has captured the full range of scales to within $\sim10\%$, while the temperature field has overfit to the distribution of the train simulation, which differs significantly from the test simulation.}
    \label{fig:hydro_fields}
\end{figure*}

Inspecting the direct output of our model, the hydrodynamic fields, gives more insight into the difficulty of our task. In Figure \ref{fig:hydro_fields}, we visualize slices of the density and temperature fields at the same location as in Figure \ref{fig:Lya_fields}, comparing our reconstructions against the ground truth fields. While there is good agreement by eye, the zoomed-in panels show discrepancies between the true and generated fields, particularly in how sharply resolved the edges of filament structures are. There are also some small-scale artifacts, especially in the temperature field, which possibly contribute to the excess of power seen at small scales in the \Lya{} $P(k)$. 

In the lower left panels, we also plot the PDFs of the two fields, computed over the entire 1024$^3$ volume of the test simulation. The full-volume PDF for the density field actually shows remarkable agreement between our model's reconstruction and the true field, with the full range of scales captured to within $\sim10\%$. This is in contrast to FGPA, which completely fails in estimating densities beyond $(\rho/\bar{\rho}) \gtrsim 10^3$.

In the temperature field, the reconstructions from FGPA and our model both differ significantly from the truth. For reference, we have also plotted the full-volume PDF of the temperature field from the training simulation, and we see that FGPA and our model both closely track the train distribution for $T/K<10^5$, indicating overfitting. However, there is a clear distributional difference between the train and test simulations, with the latter having almost an order of magnitude higher contribution from temperatures $T/K >\sim10^5$ and a subsequent deficit in the PDF at lower temperatures. This suggests we are in the low-data regime by training on a single 20 Mpc/$h$ at $\sim20$ kpc resolution, as the train and test data are not identically distributed despite having identical cosmologies and physical models. Compounding this train-test discrepancy, the clear lack of high temperatures above $T/K = 10^5$ confirms that our model is failing to correctly resolve the sharp boundaries of high-temperature shocks forming at the edges of filament structures.


\begin{figure*}
    \centering
    \includegraphics[width=\linewidth]{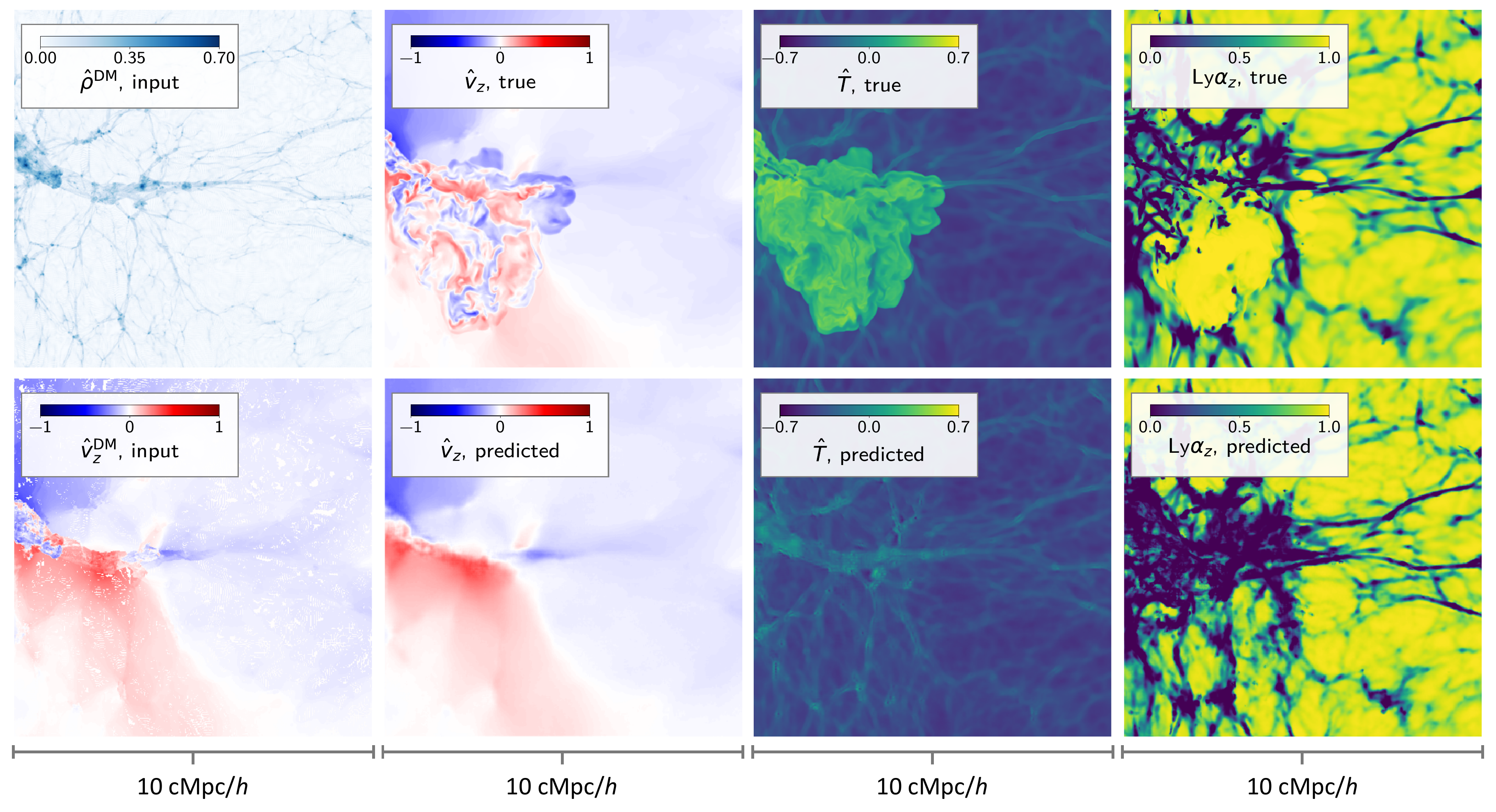}
    \caption{The widespread ($\sim5$ cMpc) shocked region in the test simulation our model fails to capture. The left column shows some of the input dark matter fields, which indicate a dense cluster but do not otherwise contain information about the exact location and shape of a shock front. The middle panels show the baryon velocity and temperature fields at the same location, and reveal that the model has failed to capture the shock features. At far right, we show the \Lya{} slices corresponding to this shocked region.}
    \label{fig:shocks}
\end{figure*}

Inspecting the test simulation more closely, we find a rare widespread shocked region of size $\sim5$ cMpc, a significant fraction of the length of the entire box. In Figure \ref{fig:shocks} we show slices of this region for the true fields and our model's reconstructions. It is clear that the model does not reproduce the shock structure, instead painting in a fairly smooth rendering of the filament structures in the area. Given that the model is trained on randomly-cropped sub-volumes of size 2.5 cMpc, it is unreasonable to expect coherent shock structures as large as this to be well-represented. Furthermore, as such structures form $O(1)$ times per 20 Mpc/$h$ box, they are exceedingly rare in the training data. In fact, there is no shocked region as large as this in the training simulation, hence the clear difference between the temperature PDFs in the train and test simulations.

\section{Conclusion}
\label{sec:Conclusion}

In this work we have presented a convolutional neural network approach to map from collisionless n-body simulations to fully hydrodynamical simulations, reconstructing the baryon density, temperature and velocity fields. We find this mapping can match the statistical properties of the true hydrodynamical simulations across a wide range of scales. When computing an observable quantity, \lya\ flux, from our model output, we can outperform existing semi-analytical methods across all scales and match the true simulated power-spectra within $\sim 5\%$ up to $k \sim 10$ $h$/Mpc. 

The range of scales accurately mapped with this method is well beyond the resolution of next generation observations \citep{Walther2021}. This method provides an alternative to running costly hydrodynamical simulations for mock catalog generation and the construction of \lya\ power-spectra emulators. The network can conveniently be trained on small simulated boxes at a given point in cosmological and astrophysical parameter space, then applied to a large dark matter only box at inference time. Thus, as the size of the inference box grows, this approach enables increasingly impressive reductions in computational cost. 

Beyond \lya\ forest statistics, there are some notable limitations to this method. Processes not well described by the dark matter distribution are difficult to predict, particularly the high temperature gas shocks seen in Figure \ref{fig:shocks}. While these regions have limited effect on summary statistics used for \Lya\ cosmological analysis, they are of significant astrophysical interest as they are associated with the physics of the Warm Hot Intergalactic Medium (WHIM) \citep{1999ApJ...514....1C,2006MNRAS.367..113P,2019A&A...627A...5V}. Considering our neural network design, the predictions in Figure \ref{fig:shocks} are not entirely unreasonable given the input dark matter fields from the n-body simulation (visualized in the far left panels). These inputs contain a dense cluster towards the edge, but do not have obvious indicators of the scale or detailed shape of the shock front. Thus, passing these inputs to a deterministic model such as ours yields a more conservative prediction, as the model does not have the expressive capacity to represent the many possible shapes of shocked gas. We thus propose that a variational model producing multi-modal outputs would best resolve shocks like the one shown in Figure \ref{fig:shocks}. Such a model would synthesize multiple realizations of the hydrodynamic variables via some latent variable, conditioned on the input n-body fields. We explore this challenge in Horowitz et al. (2021), a companion work to this paper.

\section*{Acknowledgments}
BH is supported by the AI Accelerator program of the Schmidt Futures Foundation.
This work was partially supported by the DOE's Office of Advanced Scientific Computing Research and Office of High Energy Physics through the Scientific Discovery through Advanced Computing (SciDAC) program.
This research used resources of the National Energy Research Scientific Computing Center, a DOE Office of Science User Facility supported by the Office of Science of the U.S. Department of Energy under Contract No. DEC02-05CH11231.

\bibliography{refs}



\end{document}